\documentclass{IEEEcsmag}

\usepackage[colorlinks,urlcolor=blue,linkcolor=blue,citecolor=blue]{hyperref}
\expandafter\def\expandafter\UrlBreaks\expandafter{\UrlBreaks\do\/\do\*\do\-\do\~\do\'\do\"\do\-}
\usepackage{upmath,color}

\jvol{XX}
\jnum{XX}
\paper{8}
\jmonth{Month}
\jname{Publication Name}
\jtitle{Publication Title}
\pubyear{2021}

\usepackage{url}
\usepackage{graphicx}%
\usepackage{multirow}%
\usepackage{amsmath,amssymb,amsfonts}%
\usepackage{amsthm}%
\usepackage{mathrsfs}%
\usepackage{xcolor}%
\usepackage{textcomp}%
\usepackage{manyfoot}%
\usepackage{booktabs}%
\usepackage{algorithm}%
\usepackage{algorithmicx}%
\usepackage{algpseudocode}%
\usepackage{listings}%
\usepackage{lmodern}

\usepackage{subcaption}

\newcommand{\update}[1]{\textcolor{black}{#1}}

\begin{document}

\title{Ten Simple Rules for Catalyzing Collaborations and Building Bridges between Research Software Engineers and Software Engineering Researchers}

\author{Nasir U. Eisty}
\affil{University of Tennessee, Knoxville, TN, 37916, USA}

\author{Jeffrey C. Carver}
\affil{University of Alabama, Tuscaloosa, AL, 35487, USA}

\author{Johanna Cohoon}
\affil{{}Lawrence Berkeley National Lab, Berkeley, CA, 94720, USA}

\author{Ian A. Cosden}
\affil{Princeton University, Princeton, NJ, 08544, USA}

\author{Carole Goble}
\affil{University of Manchester, Manchester, M13 9PL, UK}

\author{Samuel Grayson}
\affil{University of Illinois, Urbana, IL, 61801, USA}

\begin{abstract}
In the evolving landscape of scientific and scholarly research, effective collaboration between Research Software Engineers (RSEs) and Software Engineering Researchers (SERs) is pivotal for advancing innovation and ensuring the integrity of computational methodologies. 
This paper presents ten strategic guidelines aimed at fostering productive partnerships between these two distinct yet complementary communities. 
The guidelines emphasize the importance of recognizing and respecting the cultural and operational differences between RSEs and SERs, proactively initiating and nurturing collaborations, and engaging within each other's professional environments. 
They advocate for identifying shared challenges, maintaining openness to emerging problems, ensuring mutual benefits, and serving as advocates for one another. 
Additionally, the guidelines highlight the necessity of vigilance in monitoring collaboration dynamics, securing institutional support, and defining clear, shared objectives. 
By adhering to these principles, RSEs and SERs can build synergistic relationships that enhance the quality and impact of research outcomes. 
\end{abstract}

\maketitle

\section{Introduction}
\label{sec_introduction}


Recognizing that the software engineering researcher (SER) and the research software engineer (RSE) communities have developed independently, it is a cross-disciplinary and cross-cultural endeavor to build bridges between them.
Currently, there is insufficient transfer of state-of-the-art knowledge between the SER and RSE communities. 
There are benefits to bridging the chasm between SERs, who primarily focus on industry applications, and RSEs, who develop code for research purposes and may not have formal training in software engineering. 
By fostering collaboration, SERs can discover novel research questions from RSE experiences, and RSEs can enhance their productivity by applying approaches and tools developed by SERs.

To facilitate this process, we have developed a guide of 10 simple rules for catalyzing collaborations and building bridges between RSEs and SERs.
The group of authors represent both of these communities and are informed by a recent Dagstuhl workshop focused on the intersection of these two communities (see Acknowledgments).
This guide outlines potential benefits and provides a straightforward set of rules to help both communities thrive in new and mutually beneficial collaborations.

Collaborations between SERs and RSEs present a vital and impactful opportunity to advance both scholarly research and software development practices~\cite{8704922, 10.1145/3084225}.
SERs can contribute \update{conceptual} insights and innovative methodologies, while RSEs can offer practical experience in developing and maintaining research software. 
By working together, they can address complex challenges more effectively, enhance the quality and reproducibility of research outputs, and accelerate the translation of research findings into practical applications~\cite{183756}. 
Such partnerships also facilitate the exchange of knowledge and skills, fostering a more integrated and efficient approach to tackling interdisciplinary problems~\cite{hansson2024measuring}. 

Achieving collaboration between SERs and RSEs requires intentional effort and the application of change theories~\cite{feitosa2023understanding}. 
This involves leveraging motivational incentives such as open development, influence, funding, and recognition while building trust, providing resources and support, and operating with transparency~\cite{koch2023sustainable}. 
True collaboration necessitates cyclic communication, mutual benefit, and shared experiences, moving beyond mere observation or one-sided interactions~\cite{8704922}.

The ten rules acknowledge the cultural differences and other challenges that separate SERs, who usually follow traditional faculty career paths, and RSEs, who are often staff embedded within a scientific team. 
While many rules propose steps that individual SERs or RSEs may take, it is also important that organizational decision-makers and funding bodies recognize the value of and impediments to cooperation between these distinct communities. 
Software's vital role in research and the larger society demands that stakeholders across the SER, RSE, funder, and decision-maker communities all pursue the kind of innovation that can only be accomplished through collaboration. 

\section{10 Simple Rules}
\label{sec_rules}
\subsection{\textbf{\textit{Rule 1: Recognize The Two Communities Are Different}}}
RSEs and SERs might seem similar on the surface due to overlapping terminology and shared technical skills, but they are distinct communities with different cultures, priorities, and workflows. 
Understanding and respecting these differences is essential for successful collaboration.  

\textbf{Appreciate Unique Roles.} RSEs are deeply embedded in research contexts, often balancing software development with domain-specific knowledge.
They focus on creating software that meets the immediate and evolving research needs, including flexibility and experimental workflows. 
Conversely, SERs are trained to develop robust, scalable, and maintainable systems emphasizing engineering principles, often aligning with industry best practices.  

\textbf{Avoid Assumptions.} Similar terms like ``code quality,'' ``agile methods,'' or ``collaboration'' can have different connotations in these communities. 
For instance, an RSE may view code quality in terms of usability for researchers, while an SER might prioritize adherence to strict coding standards and design patterns. 
Recognizing and addressing these nuanced differences ensures better communication and reduces misunderstandings.  

\textbf{Retain Awareness of Cultural Differences.} 
\update{The research culture of RSEs often encourages innovation and experimentation, particularly in the early stages of software development where prototypes and rapid iteration are common. As research software matures, however, the focus tends to shift toward creating sustainable solutions and improving the precision and reliability of outputs, with experimentation playing a more limited role.}
In contrast, SERs often operate within environments that prioritize stability, repeatability, and rigorous testing. 
Being mindful of these contrasting approaches helps bridge gaps and fosters mutual respect.  

\textbf{Celebrate Complementary Strengths.} While the two communities have different foundations, their strengths are complementary. 
RSEs bring scientific insights and adaptability, while SERs contribute engineering expertise and software scalability. 
When harnessed effectively, this synergy can lead to transformative outcomes.  

\subsection{\textbf{\textit{Rule 2: Acknowledge Collaboration Is Not Going to Just Happen}}}

Effective collaboration between RSEs and SERs requires deliberate action and sustained effort. 
It is not a spontaneous occurrence but a process that unfolds over time.

\textbf{Initiate.} Take the first step by reaching out to potential collaborators. 
This step could involve attending interdisciplinary workshops, participating in relevant online forums, or directly contacting individuals whose work aligns with your interests. 
RSEs embedded within research teams may be physically spread across a university campus and not collocated with SERs. 
Proactive in-person and online engagement is essential to establish initial connections.

\textbf{Inquire.} Engage in open dialogues to understand each other's expertise, goals, and expectations. 
Ask questions about each other's current projects, challenges, and how your skills might complement theirs. 
This mutual understanding lays the groundwork for a productive partnership.

\textbf{Invest.} Dedicate time and resources to nurture the relationship, including regular communication, setting shared objectives, and being patient as the collaboration develops. 
Recognize that building trust and synergy does not happen overnight.
It requires consistent effort and commitment.

\subsection{\textbf{\textit{Rule 3: Define Clear Goals and Outcomes}}}
Establishing clear goals and outcomes is fundamental to successful collaborations between RSEs and SERs. 
A shared understanding of objectives ensures all parties are aligned and working towards common purposes, whether it's developing reusable software, optimizing computational workflows, or creating innovative tools.
Key steps to define clear goals and outcomes:

\textbf{Initiate Open Discussions.} Begin with transparent conversations to understand each collaborator's expectations, strengths, and desired contributions. 
This dialogue helps identify overlapping interests and potential synergies.

\textbf{Set Specific and Measurable Objectives.} Utilize frameworks like SMART (Specific, Measurable, Achievable, Relevant, Time-bound) goals to articulate clear and attainable objectives. 
For example, instead of a vague goal like ``improve software performance,'' define the goal as ``reduce software execution time by 20\% within six months.''

\textbf{Document Agreed Goals.} Create a written agreement or project charter that outlines the defined goals, expected outcomes, roles, and responsibilities.
This document is a reference point throughout the collaboration, ensuring accountability and clarity.

\textbf{Establish Regular Check-ins.} Schedule periodic meetings to assess progress toward the set goals, address any challenges, and make necessary adjustments.
Regular check-ins facilitate continuous alignment and keep the collaboration on track.

\textbf{Be Flexible and Adaptive.} Recognize that research and development projects can evolve. 
Be open to refining goals as new insights emerge, ensuring the collaboration remains relevant and productive.

\subsection{\textbf{\textit{Rule 4: SERs Must Engage with RSEs in Their Professional Environments}}}
To foster effective collaboration between SERs and RSEs, it is crucial for SERs to actively participate in the professional settings of RSEs. 
This approach acknowledges the unique challenges RSEs face and demonstrates a commitment to understanding their work context.

\textbf{Understanding RSE Constraints.} RSEs often operate under specific limitations, including restricted funding, time constraints, and institutional obligations. 
These factors can limit their ability to attend external events or engage in activities outside their immediate responsibilities. 
By recognizing these constraints, SERs can approach collaboration with empathy and adaptability.

\textbf{Cultural and Environmental Differences.} The work environments of RSEs and SERs differ in duties, pressures, and availability. 
RSEs may have less flexibility in their schedules and more immediate research-related obligations. 
\update{While SERs may have more flexibility in some cases, their ability to engage in RSE-centric events depends on their project commitments and priorities. Nonetheless, when possible, their initiative to bridge this gap can be valuable.}

\textbf{Active Participation in RSE Spaces.} SERs should attend RSE conferences, workshops, and talks. 
This attendance not only facilitates networking but also provides insights into the challenges and priorities of RSEs.
\update{Such engagement signals mutual respect between the RSE and SER communities, demonstrating a willingness from both sides to collaborate on shared terms.}

\textbf{Building Trust and Collaboration.} By meeting RSEs in their professional environments, SERs can build trust and establish a foundation for sustained collaboration. 
This approach ensures that partnerships are grounded in mutual understanding and respect for each other's professional landscapes.

\subsection{\textbf{\textit{Rule 5: Identify the Intersection of Shared Research Software Challenges}}}

For collaborations between RSEs and SERs to be effective, it is essential to pinpoint common challenges that are both practically significant and academically valuable. 
This alignment ensures the partnership addresses real-world issues while contributing to scholarly discourse.

\textbf{SERs Must Understand RSE Needs.} RSEs often grapple with challenges such as ensuring software reproducibility, managing complex data workflows, and integrating diverse computational tools. 
Engaging in open dialogues with RSEs can reveal specific problems they encounter, providing a foundation for targeted solutions.

\textbf{Collaboration Must Align with SER Research Interests.} SERs typically focus on areas like software scalability, optimization, and the development of novel engineering methodologies, \update{as well as other critical aspects like security, data integrity, and maintainability.}
Identifying challenges that not only address RSE needs but also offer opportunities for SERs to explore these research domains can lead to mutually beneficial projects.

\textbf{Conduct Joint Requirements Gathering.} Collaboratively requirements gathering ensures that both parties clearly understand the problem space.
This process involves defining the scope, objectives, and desired outcomes, ensuring the project is practically relevant and academically rigorous.

\textbf{Focus on Publishable Outcomes.} Selecting problems that are not only significant to RSEs but also contribute to the SER community's body of knowledge increases the likelihood of producing publishable results. 
This dual focus enhances the impact of the collaboration and provides recognition for both parties.

\subsection{\textbf{\textit{Rule 6: Ensure Mutual Benefit in Collaboration}}}

For a partnership between RSEs and SERs to thrive, it is essential to create a framework where both parties gain both immediate and long-term value. 
This situation involves understanding each other's incentives, acknowledging contributions, and fostering an environment of mutual respect.

\textbf{Understand Incentives.} Recognize the motivations driving each collaborator. 
RSEs may seek to enhance research outcomes, improve software usability, or gain recognition within academic circles. 
SERs might aim to develop scalable solutions, publish findings in software engineering journals and conferences, or advance their careers through innovative projects. 
By identifying these incentives, the collaboration can be structured to address the goals of both parties.

\textbf{Explicitly Recognize Benefits.} Clearly outline the advantages each collaborator will receive. 
These benefits could include co-authorship on publications, shared intellectual property rights, or opportunities for professional development. 
Transparent discussions about expectations and rewards help prevent misunderstandings and ensure all contributors feel valued and stand to benefit.

\textbf{Short-term and Long-term Benefits.}
Understand the short-term and long-term benefits. 
Short-term benefits include improved efficiency in project execution, access to complementary expertise, and the ability to produce higher-quality software and research outputs.
Long-term benefits extend to building sustainable professional relationships, enhancing reputations within respective fields, and creating a foundation for future collaborations. 

\textbf{Promote Joint Authorship.} Encourage shared credit for collaborative work through joint authorship in publications. 
Joint authorship not only acknowledges the contributions of all parties but also enhances the credibility and reach of the research. 
Establishing clear guidelines for authorship, such as those outlined in the Contributor Roles Taxonomy (CRediT)~\cite{niso_credit}, can provide a structured approach to assigning credit based on specific contributions. 

\textbf{Foster Mutual Respect.} Avoid hierarchical dynamics where one party is viewed as subordinate. 
Instead, cultivate a partnership based on equality, where each collaborator's expertise is respected and valued. 
This approach leads to more effective problem-solving and a more harmonious working relationship.

 \subsection{\textbf{\textit{Rule 7: Maintain an Open Mind Toward Emerging Challenges}}}

While focusing on shared research software problems is essential, it is equally important to remain receptive to new challenges that may arise beyond the current scope of collaboration. 
This openness can lead to innovative solutions and expand the impact of the partnership between RSEs and SERs.

\textbf{Embrace Flexibility.} The research landscape is dynamic, with new technologies and methodologies continually emerging. 
By staying adaptable, both RSEs and SERs can pivot to address unforeseen challenges or capitalize on novel opportunities that may not have been apparent at the outset.

\textbf{Encourage Continuous Dialogue.} Regular communication fosters an environment where new ideas can surface. 
Establishing forums for brainstorming and feedback allows team members to share insights and identify potential areas for exploration beyond the initial project scope.

\textbf{Leverage Diverse Expertise.} The unique perspectives of RSEs and SERs can uncover challenges that might be overlooked within a single discipline. 
By valuing and integrating these diverse viewpoints, the collaboration can tackle a broader range of problems, leading to more comprehensive and innovative solutions.


\subsection{\textbf{\textit{Rule 8: Actively Advocate for Each Other}}}

In collaborations between RSEs and SERs, mutual advocacy is crucial for fostering understanding, recognition, and support within and across their respective communities.

\noindent \textbf{RSEs Advocating for SERs:}

\begin{itemize}
    \item \textit{Facilitate Access and Provide Case Studies:} RSEs can offer SERs access to real-world research environments and datasets, enabling them to apply their engineering expertise to meaningful problems. By sharing detailed case studies, RSEs help SERs understand the unique challenges in research settings, leading to more effective and relevant solutions.
    \item \textit{Highlight SER Contributions:} Within the research community, RSEs can showcase the impact of SERs' work on advancing research projects. These highlights can include acknowledging their role in developing robust software solutions and improving research methodologies, thereby enhancing the appreciation of engineering contributions in research.
\end{itemize}

\noindent \textbf{SERs Advocating for RSEs:}

\begin{itemize}
    \item \textit{Increase Visibility of RSE Work:} SERs can promote the successes of RSEs by highlighting their contributions in engineering forums, conferences, and publications. This advocacy brings attention to the critical role RSEs play in bridging the gap between research and software development.
    \item \textit{Support RSE Recognition:} Within engineering circles, SERs can advocate for including RSEs in discussions about software development best practices, tool development, and project management. This helps integrate RSE perspectives into broader engineering conversations.
\end{itemize}

\noindent \textbf{Advocacy Across Communities:}
\begin{itemize}
    \item \textit{Promote Cross-Community Understanding:} Both RSEs and SERs should actively share insights about each other's roles and contributions within their respective communities. This cross-pollination of knowledge fosters a culture of respect and collaboration.
    \item \textit{Champion Collaborative Successes:} Jointly celebrating successful projects and outcomes in both research and engineering forums underscores the value of interdisciplinary collaboration and sets a precedent for future partnerships.
\end{itemize}

\subsection{\textbf{\textit{Rule 9: Maintain Vigilance and Recognize When Collaborations Are Off Course}}}

Effective collaborations between RSEs and SERs require ongoing attention to ensure mutual benefit and alignment with shared goals. 
It is essential to monitor the partnership's dynamics and address any issues promptly to sustain a win-win relationship.

\noindent \textbf{Recognize Signs of Imbalance:}

\begin{itemize}
    \item \textit{Unreciprocated Contributions:} Be alert to situations where one party consistently contributes without receiving equivalent value in return. For instance, an SER might provide technical expertise without opportunities for publication, or an RSE might send data and statistics without any insight, recommendations, or collaborative input.
    \item \textit{Divergence from Shared Objectives:}  If a collaborator senses that the partnership's work has deviated from its original goals and benefits only one party, engage in an exercise to capture perspectives and surface divergence. Have each collaborator privately identify the primary work to be done, its value to the collaboration, and its value to the broader community. Present your thoughts to the group and assess together if there are major differences.
\end{itemize}

\noindent \textbf{Assess the Nature of Service Work~\cite{10.1371/journal.pcbi.1004214}:}

\begin{itemize}
    \item \textit{Evaluate Impact on Career Development:} While some tasks may be performed as favors or goodwill gestures, it is important to consider their impact on professional growth. Engaging in activities that do not contribute to career advancement can lead to dissatisfaction over time.
    \item \textit{Ensure Overall Balance:} Not every individual task needs to provide equal benefit to both parties. However, the collaboration as a whole should offer balanced advantages, ensuring that both RSEs and SERs find value in the partnership.
\end{itemize}

\noindent \textbf{Implement Regular Check-ins:}

\begin{itemize}
    \item \textit{Schedule Periodic Reviews:} Establish regular meetings to discuss the collaboration's progress, address any concerns, and realign objectives as needed.
    \item \textit{Foster Open Communication:} Encourage honest dialogue about what's working and what's not. This transparency helps identify issues early and find mutually agreeable solutions.
\end{itemize}

\noindent \textbf{Be Prepared to Adjust or Conclude Collaborations:}
\begin{itemize}
    \item \textit{Adapt Strategies:} If the partnership is not yielding the desired outcomes, be willing to modify the approach or redefine roles to better align with shared goals.
    \item \textit{Know When to Step Back:} If, after concerted efforts, the collaboration remains unbalanced or unproductive, it may be prudent to amicably conclude the partnership, preserving professional relationships for future opportunities.
\end{itemize}

\subsection{\textbf{\textit{Rule 10: Secure Institutional Support}}}
Securing institutional support is crucial for fostering effective collaborations between RSEs and SERs. Advocating for policies and funding that recognize their critical roles can lead to the creation of frameworks for collaborative positions, shared goals, and appropriate reward structures.

\textbf{Advocate for Recognition and Funding.} Institutions should acknowledge the essential contributions of RSEs and SERs to research and innovation. Advocacy efforts can focus on:
\begin{itemize}
    \item \textit{Policy Development.} Encouraging the establishment of policies that formally recognize the roles of RSEs and SERs within the academic and research framework.
    \item \textit{Dedicated Funding.} Securing funding streams specifically allocated for the recruitment, retention, and professional development of RSEs and SERs. 
For instance, the National Institutes of Health (NIH) offers the Research Software Engineer (RSE) Award to support exceptional RSEs contributing to NIH-funded research~\cite{nih_rfa_od_24_011}.
\end{itemize}

\textbf{Create Frameworks for Collaborative Positions.}
Developing structured frameworks can facilitate effective collaboration:
\begin{itemize}
    \item \textit{Joint Appointments:} Establishing positions that allow individuals to work across departments or projects, bridging the gap between research and software engineering.
    \item \textit{Interdisciplinary Teams:} Forming teams that include both RSEs and SERs to promote knowledge exchange and innovation.
    \item \textit{Performance Metrics:} Including collaborative achievements and contributions to research software development in performance evaluations.
    \item \textit{Recognition Programs:} Implementing awards or acknowledgments for successful interdisciplinary collaborations.
\end{itemize}

\section{Conclusion}
\label{sec_conclusion}
In conclusion, fostering effective collaborations between RSEs and SERs is essential for advancing both scientific research and software engineering practices. 
Recognizing the distinct cultures, priorities, and workflows of RSEs and SERs is fundamental to building mutual respect and effective communication. 

Acknowledging that partnerships require deliberate effort encourages stakeholders to actively seek and nurture collaborative opportunities. 
Participating in each other's professional settings fosters deeper understanding and integration between the communities. 
Focusing on common problems ensures collaborations are relevant and beneficial to both parties. 
Being receptive to emerging challenges and ideas can lead to innovative solutions and expanded collaboration scopes. 
Establishing clear, reciprocal advantages sustains motivation and commitment from all collaborators.
Promoting the value and contributions of collaborators within and across communities strengthens partnerships and broadens impact. 
Regularly assessing the dynamics of the partnership allows for timely adjustments and ensures ongoing alignment with shared goals. 
Advocating for policies and resources that recognize and facilitate RSE and SER roles is crucial for sustained collaboration. 
Setting specific, measurable objectives provides direction and benchmarks for success. 

By adhering to these principles, RSEs and SERs can cultivate robust, synergistic relationships that enhance the quality and impact of their work. 
Such collaborations not only bridge existing knowledge gaps but also drive innovation and excellence in both research and software engineering domains.

\section{ACKNOWLEDGMENTS}
We want to thank the organizers of Dagstuhl Seminar 24161, \textit{``Research Software Engineering: Bridging Knowledge Gaps''}, Stephan Druskat, Lars Grunske, Caroline Jay, and Daniel S. Katz, for initiating the conversation and providing a space for the development of these rules. 
We also thank all of the participants from the Dagstuhl seminar who contributed ideas through discussion during the seminar.

This document was prepared as an account of work sponsored by the United States Government. While this document is believed to contain correct information, neither the United States Government nor any agency thereof, nor the Regents of the University of California, nor any of their employees, makes any warranty, express or implied, or assumes any legal responsibility for the accuracy, completeness, or usefulness of any information, apparatus, product, or process disclosed, or represents that its use would not infringe privately owned rights. Reference herein to any specific commercial product, process, or service by its trade name, trademark, manufacturer, or otherwise, does not necessarily constitute or imply its endorsement, recommendation, or favoring by the United States Government or any agency thereof, or the Regents of the University of California. The views and opinions of authors expressed herein do not necessarily state or reflect those of the United States Government or any agency thereof or the Regents of the University of California.

This manuscript has been authored by an author at Lawrence Berkeley National Laboratory under Contract No. DE-AC02-05CH11231 with the U.S. Department of Energy. The U.S. Government retains, and the publisher, by accepting the article for publication, acknowledges, that the U.S. Government retains a non-exclusive, paid-up, irrevocable, world-wide license to publish or reproduce the published form of this manuscript, or allow others to do so, for U.S. Government purposes.
\bibliographystyle{plain}
\bibliography{sn-bibliography}

\begin{IEEEbiography}{Nasir U.~Eisty} is an Assistant Professor of Computer Science at the University of Tennessee, Knoxville. His research interests lie in the areas of Software Engineering, AI for Software Engineering, Research Software Engineering, and Software Security. Eisty received his Ph.D degree in Computer Science from the University of Alabama. Contact him at neisty@utk.edu
\end{IEEEbiography}

\begin{IEEEbiography}{Jeffrey C. Carver} is a James R. Cudworth Professor of Computer Science at the University of Alabama. 
Carver received his Ph.D. in Computer Science from the University of Maryland.
His research interests are in Software Engineering for Research Software, Empirical Software Engineering, and Human Factors in Software Engineering.
Contact him at carver@cs.ua.edu.
\end{IEEEbiography}

\begin{IEEEbiography}{Johanna Cohoon} is a User Experience Researcher in the Scientific Data Division at Lawrence Berkeley National Lab. Cohoon received her Ph.D. in Information Science from the University of Texas at Austin. She works to facilitate high quality science through thoughtful infrastructure. Contact her at hcohoon@lbl.gov.
\end{IEEEbiography}

\begin{IEEEbiography}{Ian A. Cosden} is senior director for research software engineering at Princeton University, Princeton, NJ, 08544, USA. Cosden received his Ph.D. degree in mechanical engineering from the University of Pennsylvania. His research interests include Research Software Engineering (RSE) training and education, RSE business and functional models, and career pathways and entry points for research software engineers. Contact him at icosden@princeton.edu.
\end{IEEEbiography}

\begin{IEEEbiography}
{Carole Goble} CBE FREng is a Professor of Computer Science with 30+ years of experience in knowledge and computational analysis infrastructure for science disciplines, notably in the Life Sciences and Biodiversity. Carole is a co-founder of the UK’s Software Sustainability Institute. She is the joint head of the UK node ELIXIR, the European Research Infrastructure for Life Science, and a partner in the European Virtual Institute for Research Software Excellence (EVERSE).  Contact her at carole.goble@manchester.ac.uk
\end{IEEEbiography}

\begin{IEEEbiography}{Samuel Grayson} is a PhD candidate at the University of Illinois Urbana-Champaign. Sam studies reproducibility of research and attempts to apply his research into practice. Contact him at grayson5@illinois.edu.
\end{IEEEbiography}
\end{document}